\documentclass{article}
\usepackage{amssymb,latexsym}
\usepackage{amsmath, amsthm}
\begin{document}
\newtheorem{Lemma}{Lemma}
\newtheorem{Theorem}{Theorem}
\newtheorem{Proposition}{Proposition}
\newtheorem{theo}{Proposition}
\newcommand{\iv}{\raisebox{-1.9pt}{$\lrcorner$}\hspace*{-3.2pt}\rfloor}
\newcommand{\sgn}{\mathop {\rm sgn}\nolimits}
\newcommand{\ind}{\mathop {\rm ind}\nolimits }
\newcommand{\tr}{\mathop {\rm tr}\nolimits }
\newcommand{\be}{\begin{equation}}
\newcommand{\ee}{\end{equation}}
\newcommand{\fourg}{{}^4g}
\newcommand{\rd}{{\rm d}}
\newcommand{\Ol}{O_{\ln^*x }}

\title{Geometry of null hypersurfaces}

\author{\it Jacek Jezierski$^1$\thanks{supported by the Polish Research
Council grant KBN 2 P03B 073 24}  \\
\rm $^1$Department of Mathematical Methods in Physics, \\ University of Warsaw,
ul. Ho\.za 69, 00-682 Warszawa, Poland\\
E-mail: Jacek.Jezierski@fuw.edu.pl}

\maketitle

\abstract{%
We review some basic natural geometric objects on null hypersurfaces.
Gauss-Codazzi constraints are given in terms of
the analog of canonical ADM momentum which is a well defined tensor density
on the null surface. Bondi cones are analyzed with the help of this object.
}

\section{Introduction}  \label{intro}
In Synge's festshrift volume \cite{RP}
Roger Penrose distinguished three basic structures which a null hypersurface $N$
in four-dimensional spacetime $M$ acquires from the ambient Lorentzian geometry:
\begin{itemize}
\item the degenerate metric $\left.g\right|_N$ (see \cite{nurek} for
Cartan's classification of them and the solution of the local equivalence problem)
\item the concept of an affine parameter along each of the null geodesics from
 the two-parameter family ruling $N$
\item the concept of parallel transport for tangent vectors along each of the
      null geodesics
\end{itemize}
Using all three concepts on $N$ one can define several natural geometric objects
which we shall review in this article.

In Section 2 we remind the structures which are presented in \cite{Galloway}.
In the next section we give solutions, which are mostly based on \cite{ROMP46},
 to the following questions:
\begin{itemize}
\item What is the analog of canonical ADM momentum for the null surface?
\item What are the ''initial value constraints''?
\item Are they intrinsic objects?
\end{itemize}
 More precisely,
we remind the construction of external geometry in terms
of tensor density which is a well defined intrinsic object on a null
surface.
 We
already developed some applications of these object to the
following subjects:
\begin{itemize}
\item Dynamics of the light-like matter shell from matter Lagrangian which is
   an invariant scalar density on $N$ \cite{PRD65}
\item Dynamics of gravitational field in a finite volume with
null boundary and its application to black holes thermodynamics
\cite{ECPhD} (see also in this volume)
\item Geometry of crossing null shells \cite{JMP44}.
\end{itemize}

In the last section we apply our construction to Bondi cones.

\section{Natural geometric structures on $TN/K$}
We remind some standard constructions on null hypersurfaces (see \cite{Galloway}):
\begin{itemize}
\item time-oriented Lorentzian manifold $M$ with signature $(-,+,+,+)$.
\item null hypersurface $N$ -- submanifold with codim=1 and degenerate induced metric
 $\left.g\right|_N$ $(0,+,+)$, $K$ -- time-oriented non-vanishing null vector field such
 that $K_p^\perp = T_p N$ at each point $p\in N$
 \begin{enumerate}
\item $K$ is null and tangent to $N$, $g(X,K)$=0 iff $X$ is a vector field tangent to $N$
\item integral curves of $K$ are null geodesic generators of $N$
\item $K$ is determined by $N$ up to scaling factor being any positive function.
\end{enumerate}
\item $T_p N/K:= \left\{ {\overline X} : X\in T_p N \right\}$ where ${\overline X}=[X]_{{\rm mod}\; K}$
is an equivalence class of the relation ${\rm mod}\; K$ defined as follows:
\[ X\equiv Y ({\rm mod}\; K) \iff X-Y \; \mbox{is parallel to} \; K . \]
\item $\displaystyle T N/K:= \cup_{p\in N}T_p N/K$ vector bundle over N with 2-dimensional
fibers (equipped with Riemannian metric $h$),
the structure does not depend on the choice of $K$ (scaling factor)
 \[ h : T_p N/K \times T_p N/K \longrightarrow {\mathbb R} \, , \; \quad
        h ({\overline X}, {\overline Y})= g(X,Y) \; . \]
\end{itemize}

\begin{itemize}
\item null Weingarten map $b_K$ (depending on the choice of scaling factor,
 in non-degenerate
case one can always take unit normal to the hypersurface but in null case
 the vectorfield $K$ is no longer transversal to $N$ and has always scaling
 factor freedom because its length vanishes)
\[  b_K : T_p N/K \longrightarrow  T_p N/K  \, , \; \quad
        b_K({\overline X})= \overline{ \nabla_X K} \, ;\] \[
        b_{fK}=fb_K \, , \quad f\in C^{\infty}(N) \, , \quad f>0 \, .\]
\item null second fundamental form $B_K$ (bilinear form associated to $b_K$ via $h$)
  \[  B_K : T_p N/K \times T_p N/K \longrightarrow {\mathbb R}  \] \[
        B_K({\overline X}, {\overline Y})= h(b_K({\overline X}), {\overline Y})
        = g( \nabla_X K, Y)\]
  Moreover, $b_K$ is self-adjoint with respect to $h$ and $B_K$ is symmetric.
 \item $N$ is totally geodesic (i.e. restriction to $N$ of the Levi-Civita connection
 of $M$  is an affine connection on $N$, any geodesic in $M$ starting tangent to $N$
 stays in $N$) $\iff$ $B=0$ (non-expanding horizon is a typical example).
\item null mean curvature of $N$ with respect to $K$
\[ \theta := \tr b = \sum_{i=1}^2 B_K({\overline e_i},{\overline e_i})=
\sum_{i=1}^2 g( \nabla_{e_i} K, e_i)\]

where ${\overline e_i}$ is an orthonormal basis for $T_p N/K$,
$e_i$ is an orthonormal basis for $T_p S$ in the induced metric
on $S$ which is a two-dimensional submanifold of $N$ transverse to $K$.

\end{itemize}

We assume now that $K$ is a geodesic vector field i.e. $\nabla_K K=0$.
Let us denote by prime covariant differentiation in the null direction:
\[ {\overline Y}':=\overline{\nabla_K Y}\, , \quad
b'({\overline Y}):=b({\overline Y})'-b({\overline Y}') \] 
From Riemann tensor we build the following curvature endomorphism
\[ R: T_p N/K \longrightarrow  T_p N/K \, , \quad R({\overline X})=\overline{Riemann(X,K)K} \]
and we get a Ricatti equation
\be\label{Rb} b' + b^2 + R =0 \, . \ee
Taking the trace of (\ref{Rb}) we obtain well-known Raychaudhuri equation:
\be\label{Re} \theta'=-Ricci(K,K) -B^2 \, , \quad B^2=\sigma^2+\frac12\theta^2 \ee
where $\sigma$ is a shear scalar corresponding to the trace free part of $B$.
A standard application of the Raychaudhuri equation gives the following
\begin{Proposition}
Let $M$ be a spacetime which obeys the null energy condition, i.e. Ricci$(X,X)\geq 0$
for all null vectors $X$, and let $N$ be a smooth null hypersurface in $M$.
If the null generators of $N$ are future geodesically complete then $N$ has nonnegative
null mean curvature i.e. $\theta \geq 0$.
\end{Proposition}

\section{Canonical momentum on null surface}


For non-degenerate hypersurface we define the canonical ADM momentum:
\be
{P}^{kl} := \sqrt{\det g_{mn}} ({g}^{kl} g_{ij}{\cal K}^{ij} -  {\cal K}^{kl})\, ,
\ee
where ${\cal K}^{kl}$ is the second fundamental form (external curvature) of the
imbedding of the hypersurface into the spacetime $M$.

Gauss-Codazzi equations for non-degenerate hypersurface are the following:
\begin{equation}
     P_i{^l}{_{| l}} = \sqrt{\det g_{mn}}\, G_{i\mu}n^{\mu}\quad
     (=8\pi\sqrt{\det g_{mn}}\, T_{i\mu}n^{\mu})\, ,
\end{equation}
\be
(\det g_{mn}){\cal R}- {P}^{kl}{P}_{kl}
 + \frac{1}{2} ({P}^{kl}{g}_{kl})^2  = 2(\det g_{mn}) G_{\mu \nu}n^{\mu}n^{\nu}\ee
 \begin{equation}\nonumber
 (=16\pi(\det g_{mn}) T_{\mu \nu}n^{\mu}n^{\nu}) \, ,
\end{equation}
where
$\cal R$ is the (three--dimensional) scalar curvature of $g_{kl}$,
$n^\mu$ is a four--vector normal to the hypersurface,
  $T_{\mu\nu}$ is an energy--momentum tensor of the matter field,
and the calculations have been made with respect to the non-degenerate
induced three--metric $g_{kl}$
("$|$" denotes covariant derivative, indices are raised and lowered with respect to that metric etc.)

A null hypersurface in a Lorentzian spacetime $M$ is a
three-dimensional submanifold $N \subset M$ such that the
restriction $g_{ab}$ of the spacetime metric $g_{\mu\nu}$ to $N$
is degenerate.

We shall often use adapted coordinates, where coordinate $x^3$ is
constant on $N$. Space coordinates will be labeled by $k,l =
1,2,3$; coordinates on $N$ will be labeled by $a,b=0,1,2$;
finally, coordinates on $S$ will be labeled by $A,B=1,2$.
Spacetime coordinates will be labeled by Greek characters
$\alpha, \beta, \mu, \nu$.

We will show in the sequel that null-like counterpart of initial
data $(g_{kl}, P^k{_l})$ consists of the metric $g_{ab}$ and
tensor density $Q^a{_b}$ which is
a mixed (contravariant-covariant) tensor density.

The non-degeneracy of the spacetime metric implies that the
metric $g_{ab}$ induced on $N$ from the spacetime metric
$g_{\mu\nu}$ has signature $(0,+,+)$. This means that there is a
non-vanishing null-like vector field $K^a$ on $N$, such that its
four-dimensional embedding $K^\mu$ to $M$ (in adapted coordinates
$K^3=0$) is orthogonal to $N$. Hence, the covector $K_\nu = K^\mu
g_{\mu\nu} = K^a g_{a\nu}$ vanishes on vectors tangent to $N$ and,
therefore, the following identity holds:
\begin{equation}\label{degeneracy}
  K^a g_{ab} \equiv 0 \ .
\end{equation}
It is easy to prove that integral curves of
$K^a$, after a suitable reparameterization, are geodesic curves of
the spacetime metric $g_{\mu\nu}$. Moreover, any null
hypersurface $N$ may always be embedded in a one-parameter
congruence of null hypersurfaces.

We assume that topologically we have $N = {\mathbb R}^1 \times
S^2$. Since our considerations are purely local, we fix the
orientation of the ${\mathbb R}^1$ component and assume that
null-like vectors $K$ describing degeneracy of the metric $g_{ab}$ of
$N$ will be always compatible with this orientation. Moreover, we
shall always use coordinates such that the coordinate $x^0$
increases in the direction of $K$, i.e. ~inequality $K(x^0) = K^0
> 0$ holds. In these coordinates degeneracy fields are of the form
$K = f(\partial_0-n^A\partial_A)$, where $f > 0$, $n_A = g_{0A}$
and we rise indices with the help of the two-dimensional matrix
${\tilde{\tilde g}}^{AB}$, inverse to $g_{AB}$.

If by $\lambda$ we denote the two-dimensional volume form on each
surface $\{x^0 = \mbox{\rm const.}\}$:
\begin{equation}\label{lambda}
  \lambda:=\sqrt{\det g_{AB}} \ ,
\end{equation}
then for any degeneracy field $K$ of $g_{ab}$ the following
object
\[
v_{K} := \frac {\lambda}{K(x^0)}
\]
is a well defined scalar density on $N$.  This means
that \[ {\bf v}_K := v_K \rd x^0 \wedge \rd x^1 \wedge \rd x^2\]
 is a
coordinate-independent differential three-form on $N$. However,
$v_K$ depends upon the choice of the field  $K$.

It follows immediately from the above definition that the
following object:
\[
\Lambda = v_K \ K
\]
is a well defined (i.e. ~coordinate-independent) vector density on
$N$.

Obviously, it {\em does not depend} upon any choice of the
field $K$:
\begin{equation}\label{Lambda}
\Lambda =  \lambda (\partial_0-n^A\partial_A)
\end{equation}
and it is an intrinsic property of the internal geometry
$g_{ab}$ of $N$. The same is true for the divergence $\partial_a
\Lambda^a$ which is, therefore, an invariant, $K$-independent,
scalar density on $N$. Mathematically (in terms of differential
forms) the quantity $\Lambda$ represents the two-form:
\[
{\bf L} := \Lambda^a \left( \partial_a \; \iv  \;  \rd x^0 \wedge
\rd x^1 \wedge \rd x^2 \right) \, ,
\]
whereas the divergence represents its exterior derivative (a
three-from): d${\bf L} := \left( \partial_a \Lambda^a \right)\rd x^0
\wedge \rd x^1 \wedge \rd x^2$.

In particular, a null surface with
vanishing d${\bf L}$ is the {\em non-expanding horizon}.

Both objects ${\bf L}$ and ${\bf v}_K$ may be defined
geometrically, without any use of coordinates. For this purpose we
note that at each point $p \in N$ the tangent space $T_p N$ may be
quotiented with respect to the degeneracy subspace spanned by $K$.
The quotient space $T_p N/K$ carries a non-degenerate Riemannian metric $h$ and,
therefore, is equipped with a volume form $\omega$ (its coordinate
expression would be: $\omega = \lambda \ \rd x^1 \wedge \rd x^2$).

The
two-form ${\bf L}$ is equal to the pull-back of $\omega$ from the
quotient space $T_p N/K$ to $T_p N$
\[ \pi : T_p N \longrightarrow T_p N/K  \, , \quad {\bf L}:=\pi^*\omega \, . \]

The three-form ${\bf v}_K$ may be
defined as a product: \[ {\bf v}_K = \alpha \wedge {\bf L} \, ,\]
 where
$\alpha$ is {\em any} one-form on $N$, such that $<K,\alpha>
\equiv 1$.

We have
\[
{\rm d}{\bf L}=\theta {\bf v}_K \]
where $\theta$ is a null mean curvature of $N$.

The degenerate metric $g_{ab}$ on $N$ does not allow to define
{\em via} the compatibility condition $\nabla g = 0$, any natural
connection, which could be applied to generic tensor fields on $N$.
Nevertheless, there is one exception:
the degenerate metric defines {\em uniquely} a certain covariant,
first order differential operator.
 The operator may be applied only to mixed
(contravariant-covariant) tensor density fields ${\bf H}^{a}{_b}$,
satisfying the following algebraic identities:
\begin{eqnarray}
{\bf H}^{a}{_b} K^b = 0 \ , \label{G-1} \quad {\bf H}_{ab} = {\bf
H}_{ba} \ , 
\end{eqnarray}
where ${\bf H}_{ab} := g_{ac} {\bf H}^{c}{_b}$. Its definition
cannot be extended to other tensorial fields on $N$. Fortunately,
 the extrinsic curvature of a null-like surface and
the energy-momentum tensor of a null-like shell are described by
tensor densities of this type.

The operator, which we denote by ${\overline{\nabla} }_a$,
is defined by means of the four-dimensional
metric connection in the ambient spacetime $M$ in the following
way:\\
 Given ${\bf H}^{a}_{\ b}$, take any its extension ${\bf
H}^{\mu\nu}$  to a four-dimensional, symmetric tensor density,
``orthogonal'' to $N$, i.e. satisfying ${\bf H}^{\perp\nu}=0$
(``$\perp$'' denotes the component transversal to $N$). Define
${\overline{\nabla} }_a {\bf H}^{a}_{\ b}$ as the restriction to
$N$ of the four-dimensional covariant divergence ${\nabla}_\mu
{\bf H}^{\mu}_{\ \nu}$.
The ambiguities,
which arise when extending three-dimensional object ${\bf
H}^{a}_{\ b}$ living on $N$ to the four-dimensional one, cancel
finally and the result is unambiguously defined as a
covector density on $N$. It turns out, however, that this result
does not depend upon the spacetime geometry and may be defined
intrinsically on $N$ as follows:
\begin{equation}
\label{covariant-deg}
    {\nabla}_a {\bf H}^{a}_{\ b}  =
     \partial_a {\bf H}^{a}_{\ b} -
    \frac 12 {\bf H}^{ac} g_{ac , b} \ ,
\end{equation}
where $g_{ac , b} := \partial_b g_{ac}$,
 a tensor density ${\bf
H}^{a}_{\ b}$ satisfies identities (\ref{G-1}),
and moreover, ${\bf H}^{ac}$ is {\em any} symmetric tensor
density, which reproduces ${\bf H}^{a}_{\ b}$ when lowering an
index:
\begin{equation}\label{G-mixed}
  {\bf H}^{a}_{\ b}  = {\bf H}^{ac} g_{cb} \, .
\end{equation}
It is easily seen, that such a tensor density always exists due to
identities (\ref{G-1}), but the reconstruction of
${\bf H}^{ac}$ from ${\bf H}^{a}_{\ b}$ is not unique because
${\bf H}^{ac} + C K^a K^c$ also satisfies (\ref{G-mixed}) if
${\bf H}^{ac}$ does. Conversely, two such symmetric tensors ${\bf
H}^{ac}$ satisfying (\ref{G-mixed}) may differ only by $C K^a K^c$.
 Fortunately, this non-uniqueness does not influence the value of
(\ref{covariant-deg}).

Hence, the following definition makes sense:
\begin{equation}\label{div-final}
  {\overline{\nabla}}_a {\bf H}^{a}{_b}  :=
\partial_a {\bf H}^{a}_{\ b} - \frac 12 {\bf H}^{ac}
g_{ac , b} \ .
\end{equation}
The right-hand-side does not depend upon any choice of coordinates
(i.e. it transforms like a genuine covector density under change of
coordinates).

To express directly the result in terms of the original tensor
density ${\bf H}^{a}{_b}$, we observe that it has five
independent components and may be uniquely reconstructed from
${\bf H}^{0}{_A}$ (2 independent components) and the symmetric
two-dimensional matrix ${\bf H}_{AB}$ (3 independent components).
Indeed, identities (\ref{G-1}) may be rewritten as
follows:
\begin{align}
{\bf H}^{A}_{\ B} & =  {\tilde{\tilde g}}^{AC}{\bf H}_{CB} - n^A
{\bf H}^{0}_{\ B} \ , \label{AB}
\\ {\bf H}^{0}_{\ 0} & =  {\bf H}^{0}_{\ A} n^A \ ,  \label{00}
\\ {\bf H}^{B}_{\ 0} & =  \left( {\tilde{\tilde g}}^{BC}{\bf H}_{CA}
- n^B {\bf H}^{0}_{\ A} \right) n^A \label{B0} \ .
\end{align}
The correspondence between ${\bf H}^{a}{_b}$ and $({\bf H}^0{_A},{\bf H}_{AB})$
is one-to-one.

To reconstruct ${\bf H}^{ab}$  from ${\bf H}^{a}{_b}$ up to an
arbitrary additive term $C K^a K^b$, take the following
 (coordinate dependent) symmetric quantity:
\begin{align}
{\bf F}^{AB} & :=  {\tilde{\tilde g}}^{AC} {\bf H}_{CD }
{\tilde{\tilde g}}^{DB} - n^A {\bf H}^{0}_{\ C } {\tilde{\tilde
g}}^{CB} - n^B {\bf H}^{0}_{\ C } {\tilde{\tilde g}}^{CA} \ ,
\\ {\bf F}^{0A} & :=  {\bf H}^{0}_{\ C } {\tilde{\tilde g}}^{CA}
=: {\bf F}^{A0} \ ,
\\ {\bf F}^{00} & :=  0 \ .
\end{align}
It is easy to observe that any ${\bf H}^{ab}$ satisfying
(\ref{G-mixed}) must be of the form:
\begin{equation}\label{reconstr}
{\bf H}^{ab} = {\bf F}^{ab} + {\bf H}^{00} K^a K^b \ .
\end{equation}
The non-uniqueness in the reconstruction of ${\bf H}^{ab}$ is,
therefore, completely described by the arbitrariness in the choice
of the value of ${\bf H}^{00}$. Using these results, we finally
obtain:
\begin{eqnarray}
{\overline{\nabla} }_a {\bf H}^{a}_{\ b} & := &
\partial_a {\bf H}^{a}_{\ b} - \frac 12 {\bf H}^{ac}
g_{ac , b} =
\partial_a {\bf H}^{a}_{\ b} - \frac 12 {\bf F}^{ac}
g_{ac , b}\nonumber
\\  & = & \hspace*{-1ex} \partial_a {\bf H}^{a}_{\ b} - \frac 12 \left(
2 {\bf H}^{0}_{\ A} \ n^A_{ \ ,b} - {\bf H}_{AC}  {\tilde{\tilde
g}}^{AC}_{\ \ \ ,b} \right) \, . \label{Coda}
\end{eqnarray}
The operator on the right-hand-side of (\ref{Coda}) is
called the (three-dimen\-sio\-nal) covariant derivative of ${\bf
H}^{a}_{\ b}$ on $N$ with respect to its degenerate metric
$g_{ab}$. It is well defined
(i.e. ~coordinate-independent) for a tensor density ${\bf
H}^{a}_{\ b}$ fulfilling conditions (\ref{G-1}).
One can also show
 that the above definition coincides with the one given in
terms of the four-dimensional metric connection and,
due to (\ref{covariant-deg}), it equals:
\begin{equation}\label{int-ext}
  \nabla_\mu {\bf H}^{\mu}_{\ b} =
  \partial_\mu {\bf H}^{\mu}_{\ b} -
    \frac 12 {\bf H}^{\mu\lambda} g_{\mu\lambda , b} =
    \partial_a {\bf H}^{a}_{\ b} -
    \frac 12 {\bf H}^{ac} g_{ac , b} \ ,
\end{equation}
hence, it coincides with ${\overline{\nabla} }_a {\bf H}^{a}{_b}$
 defined intrinsically on $N$.

To describe exterior geometry of $N$ we begin with covariant
derivatives {\em along} $N$ of the ``orthogonal vector $K$''.
Consider the tensor $\nabla_a K^\mu$. Unlike in the non-degenerate
case, there is no unique ``normalization'' of $K$ and, therefore,
such an object does depend upon a choice of the field $K$. The
length of $K$  vanishes. Hence, the tensor is
again orthogonal to $N$, i.e. ~the components corresponding to
$\mu = 3$ vanish identically in adapted coordinates. This means
that $\nabla_a K^b$ is a purely three-dimensional tensor living on
$N$. For our purposes it is useful to use the ``ADM-momentum'' version
of this object, defined in the following way:
\begin{equation}\label{Q-fund}
{Q^a}_b (K) := -s \left\{ v_K \left( \nabla_b K^a - \delta_b^a
\nabla_c K^c \right) + \delta_b^a \partial_c \Lambda^c \right\}
\, ,
\end{equation}
where $s:=\sgn g^{03}=\pm 1$. Due to the above convention, the
object ${Q^a}_b (K)$ feels only {\em external
orientation} of $N$ and does not feel any internal orientation of
the field $K$.

\underline{Remark:} If $N$ is a {\em non-expanding horizon}, the last
term in the above definition vanishes.

The last term in (\ref{Q-fund}) is $K$-independent. It has been
introduced in order to correct algebraic properties of the
quantity \[ v_K \left( \nabla_b K^a - \delta_b^a \nabla_c K^c
\right) \, .\]
 One can show that
${Q^a}_b$ satisfies identities
(\ref{G-1}) and, therefore, its covariant divergence
with respect to the degenerate metric $g_{ab}$ on $N$ is uniquely
defined. This divergence enters into the Gauss--Codazzi equations,
which
relate the divergence of $Q$ with the transversal component ${\cal
G}^{\perp}_{\ b}$ of the Einstein tensor density ${\cal G}^\mu_{\
\nu} = \sqrt{|\det g |} \left( R^\mu_{\ \nu} - \delta^\mu_\nu
\frac 12 R \right)$. The transversal component of such a
tensor density is a well defined three-dimensional object living
on $N$. In coordinate system adapted to $N$, i.e.~such that the
coordinate $x^3$ is constant on $N$, we have ${\cal G}^{\perp}_{\
b} = {\cal G}^{3}_{\ b}$. Due to the fact that ${\cal G}$ is a
tensor density, components ${\cal G}^{3}_{\ b}$ {\em do not
change} with changes of the coordinate $x^3$, provided it remains
constant on $N$. These components describe, therefore, an
intrinsic covector density living on $N$.

\begin{Proposition}
The following null-like-surface version of the Gauss--Codazzi
equation is true:
\begin{equation}\label{G-C}
    {\overline{\nabla} }_a {Q}^{a}_{\ b}(K) +s v_{K} \partial_b
    \left( \frac {\partial_c \Lambda^c}{v_{K}} \right) \equiv
    -{\cal G}^{\perp}_{\ b}  \ .
\end{equation}
 \end{Proposition} The proof is given in \cite{PRD65}.
We remind the reader that the ratio between two scalar densities:
$\partial_c \Lambda^c$ and $v_K$, is a scalar function $\theta$. Its
gradient is a covector field. Finally, multiplied by the density
$v_K$, it produces an intrinsic covector density on $N$. This
proves that also the left-hand-side is a well defined geometric
object living on $N$.

The component $K^b{\cal G}^{\perp}_{\ b}$ of the equation (\ref{G-C}) is nothing but
a densitized form of Raychaudhuri  equation (\ref{Re}) for the congruence of
null geodesics generated by the vector field $K$.

\section{Initial data on asymptotic Bondi cones}

Recall (see \cite{CJK}) that in Bondi-Sachs coordinates \( (u,x,x^{A}) \) the
space-time metric takes the form:
\be\label{gB} \fourg= -xV{\rm
e}^{2\beta}\rd u^2 + 2 {\rm e}^{2\beta} x^{-2}\rd u \rd x
   + x^{-2} h_{AB} \left(\rd x^A - U^A\rd u \right)
   \left(\rd x^B - U^B\rd u \right)\;. \ee
Let us derive explicitly canonical data $(g_{ab}, Q^a{_b})$
on null surfaces $N:=\{$$u=\;$const.$\}$ which we call {\em Bondi cones}.
The intrinsic coordinates on null surface $N$ are $x^a=(x,x^A)$.
We choose null field
\be\label{K-def} K:={\rm e}^{-2\beta}x^2\partial_x \, .\ee
The components of the degenerate metric $g_{ab}$ are as follows:
\[ g_{AB}=x^{-2}h_{AB} \, , \quad g_{xA}=0=g_{xx} \; . \]
From (\ref{gB}), (\ref{K-def}) and (\ref{Q-fund}) we obtain the following formulae:
\begin{eqnarray}
sQ^a{_x}(K) &=& 0 \\
sQ^A{_B}(K) &=& -\frac12\sin\theta \, h^{AC}\left( x^{-2}h_{CB}\right)_{,x} \\
sQ^x{_A}(K) &=& x^{-2}\sin\theta\left(\beta_{,A} +\frac12{\rm e}^{-2\beta}
h_{AB}U^B_{,x} \right)
\end{eqnarray}

 If we assume that Bondi cone data
is polyhomogeneous and conformally $C^1\times C^0$-compactifiable,
it follows that (cf. \cite{TBmass})
\[ h_{AB}={\breve{h}} _{AB}(1+\frac{x^{2}}{4}\chi
^{CD}\chi _{CD})+x\chi
_{AB}+x^2\zeta_{AB}+x^{3}\xi_{AB}+\Ol(x^{4})\, \, \, ,\] where
$\zeta_{AB}$ and $\xi_{AB}$ are polynomials in $\ln x$ with
coefficients which smoothly depend upon the $x^A$'s. By definition
of the Bondi coordinates we have $\det h = \det {\breve{h}}=\sin\theta$, which
implies
${\breve{h}}^{AB}\chi_{AB}={\breve{h}}^{AB}\zeta_{AB}=0$. Further,
 \be\label{beta}
\beta= -\frac1{32}\chi^{CD}\chi_{CD}x^2 +B x^3 + \Ol(x^4) \;,\ee
 \be\label{hU}
h_{AB}U^B = -\frac12\chi_A{^B}{_{||B}}x^2 + W_Ax^3
+\Ol(x^4)\;, \ee
where $B$ and $W_A$ are again polynomials in
$\ln x$ with smooth coefficients depending upon the $x^A$'s, while
$||$ denotes covariant differentiation with respect to the unit sphere metric
${\breve{h}}$. This leads to the following approximate formulae:
\begin{eqnarray}
sQ^A{_B}(K) &=& x^{-2}\sin\theta \left( x^{-1}\delta^A{_B} - \chi^A{_B} + O(x^2)\right) \\
sQ^x{_A}(K) &=& x^{-2}\sin\theta\left(-\frac12 x \chi_A{^B}{_{||B}} + O(x^2) \right) \\
g_{AB}&=& x^{-2}\left( {\breve{h}}_{AB}+x\chi_{AB}+O(x^2)\right)
\end{eqnarray}

It is easy to verify that the asymptotic behaviour of canonical data
$(g_{ab}, Q^a{_b})$ is determined by ``free data'' $\chi_{AB}$ which agrees
with standard Bondi-Sachs approach to the null initial value formulation.

We hope that the variational formula on a truncated cone, which is
space-like inside and light-like near Scri, (proposed in \cite{WGMPjj}) can be
formulated with the help of the object $Q^a{_b}$ for arbitrary hypersurfaces,
 i.e. without assumption that the null part of the initial surface
is a Bondi cone.


\begin{thebibliography}{99}

\bibitem{Galloway} G.J.~Galloway,
{\it Maximum Principles for null hypersurfaces and null splitting
theorems}, Ann. Henri Poincar\'e {\bf 1} (2000) pp. 543--567.
\bibitem{ROMP46} J.~Jezierski, J.~Kijowski, E.~Czuchry, {\it Geometry of
null-like surfaces in General Relativity and its application to
dynamics of gravitating matter}, Rep.  Math. Phys. {\bf 46} (2000)
pp. 399--418.
\bibitem{PRD65} J.~Jezierski, J.~Kijowski, E.~Czuchry, {\it Dynamics of a
self gravitating light-like matter shell: a gauge-invariant
Lagrangian and Hamiltonian description}, gr-qc/0110018, Phys. Rev.
D {\bf 65} (2002), p. 064036.
\bibitem{JMP44} J.~Jezierski, {\it Geometry of crossing null shells},
       J. Math. Phys. {\bf 44}, no. 2 (2003) pp.
       641--661.
\bibitem{WGMPjj} J.~Jezierski, {\it Trautman-Bondi Mass For Scalar Field And Gravity}, gr-qc/9910085,
 in Coherent States, Quantization and Gravity,
  Proceedings of the XVII-th Workshop on Geometric Methods in Physics,
  editors M. Schlichenmaier, A. Strasburger,
  S. Twareque Ali and A. Odzijewicz,
   Warsaw University Press (2001), p. 165-180
\bibitem{ECPhD} E.~Czuchry, {\it Ph.D. Thesis}, University of Warsaw 2002,
in Polish.
\bibitem{CJK} P.~Chru\'sciel, J.~Jezierski and J.~Kijowski,
{\it A Hamiltonian framework for field theories in the radiating
regime}, Springer Lecture Notes in Physics, Monographs, vol. {\bf
70} (2002).
\bibitem{TBmass} P.~Chru\'sciel, J.~Jezierski and S.~{\L}\c{e}ski,
 {\it The Trautman-Bondi mass of hyperboloidal initial data
sets},  gr-qc/0307109, in preparation
\bibitem{nurek} P.~Nurowski, D.C.~Robinson, {\it Intrinsic geometry of a null
hypersurface}, Class. Quantum Grav. {\bf 17}
(2000) pp. 4065--84.
\bibitem{RP} R.~Penrose, {\it The geometry of impulsive gravitational
waves}, in: General Relativity, ed. O'Raifeartaigh, Oxford
Clarendon 1972, pp. 101--115.

\end{thebibliography}
\end{document}